\documentclass[aps,prl,twocolumn,superscriptaddress,showpacs]{revtex4}
\usepackage{graphicx}

\input epsf.sty
\begin{document}
\title{Impurity band induced by point defects in graphene}
\author{Bor-Luen Huang}
\affiliation{Department of Physics, National Tsing Hua University,
Hsinchu 30043, Taiwan}
\author{Chung-Yu Mou}
\affiliation{Department of Physics, National Tsing Hua University,
Hsinchu 30043, Taiwan} \affiliation{Physics Division, National
Center for Theoretical Sciences, P.O.Box 2-131, Hsinchu, Taiwan}
\date{\today}
\begin{abstract}
It is pointed out that point defects on graphene are strongly
correlated and can not be treated as independent scatters. In
particular, for large on-site defect potential, it is shown that
defects induce an impurity band with density of state characterized
by the Wigner semi-circle law. We find that the impurity band
enhances conductivity to the order of $4 e^2 /h $ and explains the
absence of strong localization. Furthermore,the impurity band
supports ferromagnetism with the induced magnetic moment approaching
1$\mu_B$ per defect in the limit of infinite quasi-particle
lifetime.
\end{abstract}
 \maketitle

Recent experimental realization of single-layer
graphene\cite{Geim07,Geim}has raised much interest in studying 2D
Dirac fermions in the context of condensed matter physics. One of
graphene's peculiar properties is the anomalous electronic
properties associated with defects and disorders.  The notable
example is the observed ferromagnetic state induced by bombarding
graphene with protons\cite{Esquinazi}. Another example is the finite
conductivity at the Dirac point\cite{Tan}, indicating the absence of
strong localization in graphene.  These observations appear to
deviate from what is expected for ideal and clean graphene. Since
real graphene must involve disorders, it indicates that disorders
may play an important role in these phenomena.

Theoretically, it is known that the Dirac equation supports
localized solutions in the presence of disorders. For instance, edge
states would appear and are localized near the zig-zag
edge\cite{Mou}. These localized solutions thus provide support for
the observed magnetism in carbon nanoribbons when the Coulomb
interaction is included\cite{Hikihara}. In general, magnetism due to
defects in graphene gets widely support either by calculations based
on tight-binding models or first-principle
calculations\cite{Guinea,Yazyev1,Lehtinen}. However, results
obtained are varying and appear to be sensitive to sizes and
boundary conditions of the system\cite{Yazyev2}. It is not firmly
established that ferromagnetism can exist in graphene\cite{Yazyev1}.
On the other hand, theoretical work on the transport property of
graphene also shows varying results on conductivity while
experimental data consistently indicates that the conductivity is
larger than the value $4e^2/\pi h$ obtained by many
calculations\cite{Ting}. It is therefore important to have a more
reliable method for analyzing disordered graphene.

Experimentally, there are many possible forms of defects in
graphene\cite{Geim, Hashimoto}. However, large defects with truly
localized states interact via RKKY interaction and tend to induce
antiferromagnetism\cite{Brey}. Thus these defects do not contribute
the observed ferromagnetism. Theoretically, it is known that the
electronic state near a point defect or a disk-like defect on
graphene is semi-localized with amplitude decaying as $1/r$ when the
distance $r$ of the electron to the defect is
large\cite{Pereira,Dong}. The semi-localization implies that
disk-like defects interact directly and thus may be the source for
ferromagnetism. Furthermore, it explains the dependence on sizes and
boundary conditions for numerical results. In particular, it implies
that disk-like defects interact strongly and can not be treated
perturbatively. Hence it calls for an appropriate method to take the
semi-localization into consideration.

In this Letter, we show that the semi-localization nature of defect
states enables an impurity band form near zero energy. The existence
of impurity band explains both the absence of localization and
possible existence of ferromagnetism in graphene. Specifically, we
shall investigate an infinite graphene with randomly distributed
point defects, characterized by an on-site potential $u$. In the
large $u$ limit\cite{hydrogen}, it is shown that the electronic
spectrum due to randomly distributed point defects can be mapped
into that of a random matrix. As a result, an impurity band with
density of state characterized by the Wigner semi-circle
law\cite{wigner} forms. This impurity band appears to be observed in
previous numerical results\cite{Pereira}. Due to the appearance of
anomalous density of states near the Dirac point, we show that the
conductivity is enhanced beyond $4e^2/\pi h$ . By further including
the Coulomb interaction, it is shown that the impurity band supports
ferromagnetism with induced magnetic moment depending on
quasi-particle lifetime and defect density.

We start by considering one point defect in graphene. Let the
Hamiltonian for electrons in the $\pi$ band of an infinite graphene
be $H_0$ and the defect is located at $\vec{r} = 0$. The
wavefunction $\psi $ for an electron then satisfies
\begin{equation}
\left( H_{0} +u \delta_{\vec{r},0} \right) \psi ({\vec{r}}) = E \psi
({\vec{r}}). \label{single}
\end{equation}
The solution $\psi$ can be found by resorting to the Huygens'
principle which implies the existence of a particular solution
propagating outwards from the defect to infinity. Obviously, this
solution represents a state localized near the defect and is simply
the Green's function, $G ({\vec{r},\vec{r}', E}) $, which satisfies
$ (E- H_{0} )  G ({\vec{r},\vec{r}', E}) =
\delta_{\vec{r},\vec{r}'}$. Since both $H_0$ and $E$ are real, it is
suffice to take a particular solution
\begin{equation}
\psi_p ({\vec{r}}) = A Re \left[ G ({\vec{r},0, E}) \right],
\label{wavefunc}
\end{equation}
where $A$ is a normalization constant. Clearly, because $u
\delta_{\vec{r},0} \psi ({\vec{r}}) = u \delta_{\vec{r},0} \psi
(0)$, to satisfy Eq.(\ref{single}), one requires $A
\delta_{\vec{r},0} = u \delta_{\vec{r},0} \psi_p ({\vec{r}}) =  A u
\delta_{\vec{r},0} Re \left[ G ({0,0, E}) \right]$. Hence the
condition for the existence of the localized state at $E_0$ is
\begin{equation}
\frac{1}{u} =  Re \left[ G ({0,0, E_0}) \right]. \label{energy}
\end{equation}
The above condition can be also reached by summing the Dyson series
in which $G_u ( \vec{r}, \vec{r}',E)= G ({\vec{r},\vec{r}', E}) + G
({\vec{r},0 , E}) G ({0,\vec{r}', E})/[1/u - G (0,0, E )]$ with
$G_u$ being the Green's function in the presence of the defect.
Clearly, $E_0$ obtained in Eq.(\ref{energy}) specifies the pole in
$G_u$. In the limit $u \rightarrow \infty$, only when $Re \left[ G
({0,0, E_0}) \right]=0$ supports a localized state. For general $u$,
the intersection of the curve $Re \left[ G ({0,0, E}) \right]$ with
$1/u$ gives the energy $E_0$ of the localized state. Direct
numerical calculations show that near $E=0$, $Re \left[ G ({0,0, E
}) \right] \approx - \gamma E$. Hence for finite and large $u$, $E_0
\sim 1/u$ and in the limit $u \rightarrow \infty$, $E_0$ approaches
$0$. Here $\gamma$ characterizes the slope of the Green's function
near $E=0$. To the leading order of inverse quasi-particle lifetime,
$\delta$, one finds
\begin{equation}
\gamma = \frac{\sqrt{3}}{2} \int \frac{d^2 k }{(2\pi)^2}
\frac{1}{\delta^2 +|E_k|^2},
\end{equation}
where $\pm |E_k|$ are the energy spectra of the $\pi$ band. It shows
that $\gamma$ goes to infinity as $\delta$ approach zero.

Note that $\psi_p$ is a particular solution with the boundary
condition: $\psi \rightarrow 0$ as $r \rightarrow \infty$.
Specifically, $\psi_p ({\vec{r}})$ goes as $1/r$ when $r \rightarrow
\infty$\cite{Pereira,Dong}. Since $1/r$ is not square-integrable,
$\psi_p$ is known as a semi-localized state. The semi-localization
implies the solution to Eq.(\ref{single}) depends on the boundary
condition. In general, the localized state hybridizes with extended
states at the same energy. The degree of hybridization is determined
by the boundary condition. In the case when $\psi \rightarrow \phi_k
$ as $r \rightarrow \infty$ with $\phi_k$ being the plane wave
function with $E_k=E_0$, by appropriate superposition of $\phi_k$
and $Im \left[ G ({\vec{r},0,E}) \right]$, one obtains $\psi
(\vec{r}) = \phi_k (\vec{r}) + u G ({\vec{r},0, E_0}) \psi (0)$,
which reproduces the usual scattering solution obtained by the
Lippmann-Schwinger equation. In this case, after expressing
$\psi(0)$ by $\phi_k(0)$, it is clear that $\psi_(\vec{r})$ is still
dominated by the semi-localized part, $G ({\vec{r},0, E_0})$.

As indicated in the above, since $G ({\vec{r},0, E_0})$ dominates in
the weight even if it can hybridize with extended states, it is
therefore sufficient to consider $\psi_p$. In particular, it implies
that defects would couple strongly when many defects are in present.
The strong coupling implies that point defects can not be treated
perturbatively. Clearly, the above construction can be generalized
to the case with $N_I$ impurities located at $\vec{r}_i$ with
$i=1,2,3, \cdots,$ and $N_I$. In this case, we have the obvious
solution
\begin{equation}
\psi_p ({\vec{r}}) = \sum_{i=1}^{N_I} A^i_E Re \left[
G(\vec{r},\vec{r}_i,E) \right]. \label{wavef}
\end{equation}
As a generalization of Eq.(\ref{energy}), the energy of
semi-localized states satisfy
\begin{equation}
\left|
\begin{array}{cccccc}
1/u-g_{11} & -g_{12} & -g_{13} & \cdot  & \cdot  & -g_{1N_I} \\
-g_{21} & 1/u-g_{22} & -g_{23} & \cdot  & \cdot  & -g_{2N_I} \\
-g_{31} & -g_{32} & 1/u-g_{33} & \cdot  & \cdot  & -g_{3N_I} \\
\cdot  & \cdot  & \cdot  & \cdot  & \cdot  & \cdot  \\
\cdot  & \cdot  & \cdot  & \cdot  & \cdot  & \cdot  \\
-g_{N_I1} & -g_{N_I2} & -g_{N_I3} & \cdot  & \cdot  & 1/u-g_{N_I
N_I}
\end{array}
\right| =0, \label{eigen} \end{equation} where $g_{ij} \equiv \left[
Re G(\vec{r}_i,\vec{r}_j,E) \right] = g_{ji}$. For low density of
defects, one expects that energies of semi-localized states are near
0. Since $Re \left[ G ({0,0, E }) \right] \approx - \gamma E$ for $E
\sim 0$, in the limit of $u \rightarrow \infty$, Eq.(\ref{eigen})
reduces
\begin{equation}
\left|
\begin{array}{cccccc}
\gamma E  & -g^0_{12} & -g^0_{13} & \cdot  & \cdot  & -g^0_{1N_I} \\
-g^0_{21} & \gamma E & -g^0_{23} & \cdot  & \cdot  & -g^0_{2N_I} \\
-g^0_{31} & -g^0_{32} & \gamma E & \cdot  & \cdot  & -g^0_{3N_I} \\
\cdot  & \cdot  & \cdot  & \cdot  & \cdot  & \cdot  \\
\cdot  & \cdot  & \cdot  & \cdot  & \cdot  & \cdot  \\
-g^0_{N_I1} & -g^0_{N_I2} & -g^0_{N_I3} & \cdot  & \cdot  & \gamma E
\end{array}
\right| =0, \label{random} \end{equation} where $g^0_{ij} \equiv Re
\left[ G(\vec{r}_i,\vec{r}_j,0) \right]$. For randomly distributed
positions of defects, Eq.(\ref{random}) implies that except for the
scaling factor $\gamma$, energies of the electronic states are
exactly the eigenvalues of a symmetric random matrix. In the limit
$N_I \rightarrow \infty$, it is known that the density of $\gamma E$
follows the Wigner semi-circle law\cite{Hai}. Converting the Wigner
semi-circle law to the density of electronic states, $D(E)$, we find
\begin{equation}
D(E) = \frac{n_I \gamma}{2 \pi \Delta^2} \sqrt{4 \Delta^2 -\gamma^2
E^2},
\end{equation}
where $n_I$ is the density of defects and $\Delta^2 = N_I \langle
(g^0_{ij})^2 \rangle $ with $\langle \cdots \rangle$ denoting the
average over positions of defects. Thus the electronic states due to
defects form an impurity band with width $w$ being determined by
$\Delta$ via the relation $w=4 \Delta/ \gamma$. We note in passing
that Eq.(\ref{eigen}) can be also obtained by summing the Dyson
series for $g_{ij}$. However, the existence of the impurity band is
a non-perturbative result which can not be obtained by summing each
averaged term (averaged over the impurity potential) in the Dyson
series.

For randomly distributed defects on graphene with $N$ lattice
points, the probability for a defect being located at $\vec{r}_i$ is
$1/N$. Hence $< (g^0_{ij})^2 > =\sum_{ij} (g^0_{ij})^2 /N^2$, which
shows that $< (g^0_{ij})^2 >$ is nothing but the Fourier
transformation of $(g^0_{ij})^2$ at $\vec{k} =0$. We find $\Delta^2
= \gamma n_I /2$ and hence $w= \sqrt{8 n_I / \gamma}$, which
vanishes when $\delta$ goes to zero. Note that because $g^0_{ij}$
does not vanish only when $i$ and $j$ belong to different
sublattices, the impurity band results from the bipartite nature of
graphene.
\begin{figure}
\includegraphics[width=7.0cm]{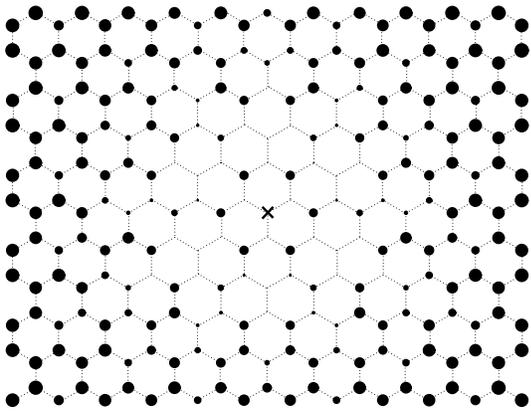}
\caption{Locations of defects that favor the spin-triplet state when
two electrons filled in semi-localized state of two defects. Here
one defect is fixed at the center, marked by X, while the other
defect is denoted by a solid dot only when the spin-triplet is
favored. Here sizes of the dots represent the strength of the
triplet state.}
\end{figure}

We first address the issue on possibility of magnetism induced by
the impurity band. We note that the impurity band is a flat-band
with hopping connecting every pair of defects. According to the
mechanism for flat-band ferromagnetism\cite{Mielke}, the impurity
band supports ferromagnetism for the case when the system can be
modeled by the Hubbard model. However, since screening is considered
to be less severe in two dimension, it calls for using more
realistic Coulomb potentials. To investigate the magnetic property
due to realistic Coulomb potential, we calculate the exchange energy
for the impurity band due to unscreened long-range Coulomb
interaction
\begin{equation}
H_C = \frac{e^2}{8 \pi \epsilon_0} \sum\limits_{i,j,\sigma,\sigma'}
C^{\dagger}_{i \sigma} C_{i \sigma} \frac{1}{|\vec{r}_i -\vec{r}_j|}
C^{\dagger}_{j \sigma'} C_{j \sigma'}.
\end{equation}
We start by considering an undoped graphene with two defects, in
which semi-localized states are filled with two electrons. In this
case, solutions to Eq.(\ref{random}) split into $E_{+}$ and $E_-$.
with the splitting, $E_+ - E_- = 2 |g^0(\vec{r}_i-\vec{r}_j)|$.
Hence the splitting oscillates in the same way as $g^0$ does and
vanishes when $i$ and $j$ belong to the same sublattice. Obviously,
two electrons can be filled in $E_{\pm}$ separately or filled in the
same states with the spin state being singlet. If the Hund's rule
dominates, two electrons fill in $E_{\pm}$ separately with their
spins being in the triplet state. Based on solutions to
Eq.(\ref{eigen}) and the corresponding wavefunctions, both Coulomb
and exchange integrals are calculated so that the preferred spin
state can be found. Fig.~1 shows our numerical calculations on
locations of defects that favor the spin-triplet state for
$u=\infty$. Here one defect is fixed at the center, marked by X,
while the other defect is shown by a solid dot only when the
spin-triplet is favored. It is seen that for large distances when
the splitting is small, spin triplet state is always favored; while
the spin singlet state is favored only for some of lattice points at
short distances when the splitting is large and exchange energy gain
does not win over the splitting.

To extend the above analysis to finite density of defects, we first
note that the normalization of the wavefunction in Eq.(\ref{wavef})
implies that $\langle \sum_{\vec{r}} \psi^2_p (\vec{r}) \rangle =1$.
If we set $E \sim 0$, we find $\langle A^i_E A^j_E \rangle =
\delta_{ij}/N_I \gamma $. By expanding $C^{\dagger}_{i \sigma} =
\sum_{E,j} A^j_E g(\vec{r}_i,\vec{r}_j,E)C^{\dagger}_{E \sigma}$ and
approximating $E$ to $0$ in $g(\vec{r}_i,\vec{r}_j,E)$, it is clear
that $\langle C^{\dagger}_{i \sigma} C_{i \sigma} \rangle$ is
determined by $\langle C^{\dagger}_{E \sigma} C_{E \sigma} \rangle$.
We find that the competition between ferromagnetic,
anti-ferromagnetic, and nonmagnetic states is determined by the
exchange energy
\begin{equation}
\langle H_C \rangle_{exchange} = -
\frac{e^2(n^2_{\uparrow}+n^2_{\downarrow})}{8 \pi \epsilon_0
\gamma^2} \sum\limits_{i,j} \frac{1}{|\vec{r}_i -\vec{r}_j|} \left(
\sum\limits^{N_I}_{k=1} g^0_{ik} g^0_{jk} \right)^2,
\label{exchange}
\end{equation}
where $n_{\sigma} = N_{\sigma} /N_I$ are fractions of electrons in
the state with the z component of spin being $\sigma$. In the
ferromagnetic state, $n_{\uparrow} \neq n_{\downarrow}$. Therefore,
the Fermi energies for different spin states are also different. Let
$E_{\sigma}$ be the corresponding Fermi energy in the spin state
$\sigma$. The total energy in the impurity band for each spin state
is $\int_{-w/2}^{E_\sigma } dE E D(E)$. For an undoped graphene,
$E_{\downarrow}=-E_{\uparrow}$. We find that $
n_{\uparrow}-n_{\downarrow}= 2( y \sqrt{1-y^2} + \sin^{-1} y) /\pi$
with $y=\gamma E_{\uparrow}/2 \Delta$, while the change of the total
energy in the impurity band per side is
\begin{equation}
\Delta k  = \frac{2n_I w}{3\pi } \left[ 1 - (1 - y^2)^{3/2} \right].
\end{equation}
Using Eq.(\ref{exchange}), the dependence of the exchange energy per
site on $n_{\uparrow}-n_{\downarrow}$ can be extracted and we find
\begin{equation}
\Delta E_{exchange}  =- \frac{e^2}{4 \pi^3 \epsilon_0 a}
(2+\sqrt{3})n^2_I \left( y \sqrt{1-y^2} + \sin^{-1} y \right)^2,
\label{Eexchange}
\end{equation}
where $a=2.46 \AA$ for graphene. In deriving Eq.(\ref{Eexchange}),
we have expressed $g^0_{ij}$ and $1/|\vec{r}_i -\vec{r}_j|$ in the
momentum space and approximated functions that are smooth in $q$ by
the corresponding values at Dirac points. It is clear that
minimizing $\Delta k + \Delta E_{exchange}$ with respect to $y$
determines the induced magnetic moment. Note that the value $y_0$
where the minimum occurs is determined by the intersection of
$y\sqrt{1-y^2}+\sin^{-1} y$ and $sy$ with $s=0.072 w(eV)/n_I$.
Fig.~2(a) shows a typical dependence of magnetic moment per defect
($m \mu_B$) on the width ($w$) of the impurity band for fixed defect
density. It is clear that increasing the bandwidth would reduce the
induced magnetic moment, which is consistent with the trend shown in
Fig.~1. In the limit of infinite quasi-particle lifetime, because
$\gamma$ goes to infinity and $w= \sqrt{8 n_I/\gamma}$, $s$
approaches zero. Hence $y_0$ approaches one so that the induced
magnetic moment approaches 1$\mu_B$ per defect, in consistent with
the first principle calculation\cite{Yazyev1}. On the contrary, when
temperature increases, $\delta$ increases and $\gamma$ decreases,
hence the induced magnetic moment decreases. For fixed
quasi-particle lifetime, because $\Delta k$ and $\Delta
E_{exchange}$ have different $n_I$ dependence: $\Delta k \sim
n^{3/2}_I$ and $\Delta E_{exchange} \sim n^2_I$, their competition
leads to a threshold in the defect density to induce ferromagnetism,
as shown in Fig.~2(b).
\begin{figure}
\rotatebox{-90}{\includegraphics*[width=7.5cm]{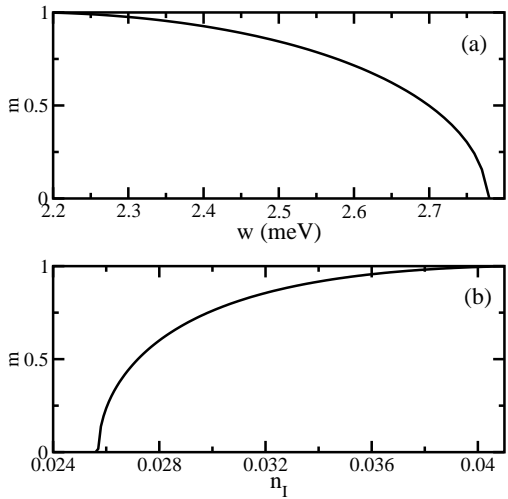}}
\caption{(a) Typical dependence of magnetic moment ($m \mu_B$) per
defect on the width ($w$) of the impurity band for $n_I=0.0001$.(b)
Typical dependence of $m$ on the defect density $n_I$ for fixed
quasi-particle lifetime. Here $\delta=0.001t$. Clearly, for fixed
quasi-particle lifetime, a threshold in defect density is necessary
to stabilize the ferromagnetic state.}
\end{figure}

We now address the issue of the transport property by explicitly
calculating the conductivity using the Kubo formula\cite{Bruus}.
Here the total current is give by $\vec{J} = i e t/\hbar \sum_{j,
\delta} \vec{\delta} C^{\dagger}_{j+\delta \sigma} C_{j \sigma}$
with $t$ being the hopping amplitude and $\vec{\delta}$ connecting
nearest neighbors. In zero temperature, we find the dc conductivity
is given by
\begin{equation}
\sigma  = 2 \pi \hbar N \left. \langle J^x_{EE'} \rangle D(E) D(E')
\right|_{E=0,E'=0},  \label{conductivity}
\end{equation}
where we have expressed $C_{i \sigma}$ in terms of $C_{E \sigma}$.
By expressing $J^x_{00}$ in the momentum space, we obtain
\begin{equation}
\sigma  = \frac{4 e^2}{h} \frac{ w^2 t^2 }{2} I , \label{cond1}
\end{equation}
where $I = \int \frac{d^2 k }{(2\pi)^2} \sin^2 (k_x/2)
|E_k|^4/(\delta^2 +|E_k|^2)^4$. For small $\delta$, because $I \sim
1/(2 \pi t^2 \delta^2)$, we find $\sigma$ is larger than the value
$4 e^2 /h$ by the factor $ w^2 /4 \pi \delta^2$. Since typical mean
free path for graphene is about 30 nanometers, $n_I \sim 10^{-5} -
10^{-6}$. Therefore, for $\delta /t \sim 10^{-3}$, we find that the
enhance factor is from the order $O(1)$, in agreement with
experimental observations\cite{Ting}.

To summarize, we show that point defects on graphene are strongly
correlated so that an impurity band near zero energy forms with the
density of state characterized by the Wigner semi-circle law. It is
shown that the impurity band supports ferromagnetism with induced
magnetic moment depending on quasi-particle lifetime and defect
density. Furthermore, we find that the induced magnetic moment
approaches 1$\mu B$ per defect in the limit of infinite
quasi-particle lifetime and a threshold in defect density is
required to stabilize ferromagnetism at fixed temperatures. Finally,
the impurity band enhances the conductivity at the Dirac point to
the order of $4 e^2 /h$, in consistent with experimental
observations.

We thank Profs. T. K. Ng, Ting-Kuo Lee, and Hsiu-Hau Lin for
discussions. This work was supported by the National Science Council
of Taiwan.


\begin{thebibliography}{99}
\bibitem{hydrogen} Defects that can be modelled by large $u$ may result from hydrogen
chemisorption or ideal point vacancies. See, for example, Ref.
\cite{Yazyev1}.
\end{thebibliography}
\end{document}